\documentclass[conference]{IEEEtran}
\IEEEoverridecommandlockouts
\usepackage[whole]{bxcjkjatype}

\usepackage{cite}
\usepackage{amsmath}
\usepackage{amssymb}
\usepackage{amsfonts}
\usepackage{algorithmic}
\usepackage{dsfont}
\usepackage{graphicx}
\usepackage{hhline}
\usepackage{multicol}
\usepackage{outlines}
\usepackage{textcomp}
\usepackage{xcolor}
\def\BibTeX{{\rm B\kern-.05em{\sc i\kern-.025em b}\kern-.08em
    T\kern-.1667em\lower.7ex\hbox{E}\kern-.125emX}}
\begin{document}

\title{Shadow Art Kanji: Inverse Rendering Application\\
\vspace{2.5mm}

\thanks{Osaka University Computer Vision Laboratory, FrontierLab@OsakaU}
}

\author{\IEEEauthorblockN{William Louis Rothman}
\IEEEauthorblockA{
\textit{College of Computing, Data Science, and Society} \\
\textit{University of California, Berkeley}\\
Berkeley, California, United States}
\and
\IEEEauthorblockN{Yasuyuki Matsushita, Ph.D.}
\IEEEauthorblockA{
\textit{Graduate School of Information Science and Technology} \\
\textit{Osaka University}\\
Suita, Osaka, Japan}}
\maketitle

\begin{abstract}
    Finding a balance between artistic beauty and machine-generated imagery is always a difficult task.
This project seeks to create 3D models that, when illuminated, cast shadows resembling Kanji characters. It aims to combine artistic expression with computational techniques, providing an accurate and efficient approach to visualizing these Japanese characters through shadows.
\end{abstract}

\section{Introduction}
\subsection{Research Background}

This project was inspired by an article written by \textit{Mitsuba: Physically Based Rendering} \cite{b1} to showcase their inverse renderer's capabilities, which details a program for optimizing a curve to project two distinct, desired shadows \cite{b2}. To do this, they use a projective sampling integrator to morph a two-dimensional manifold embedded in three-dimensional space to match the target shadows \cite{b3}.

\begin{figure}[htbp]
\centerline{\includegraphics[width=250px]{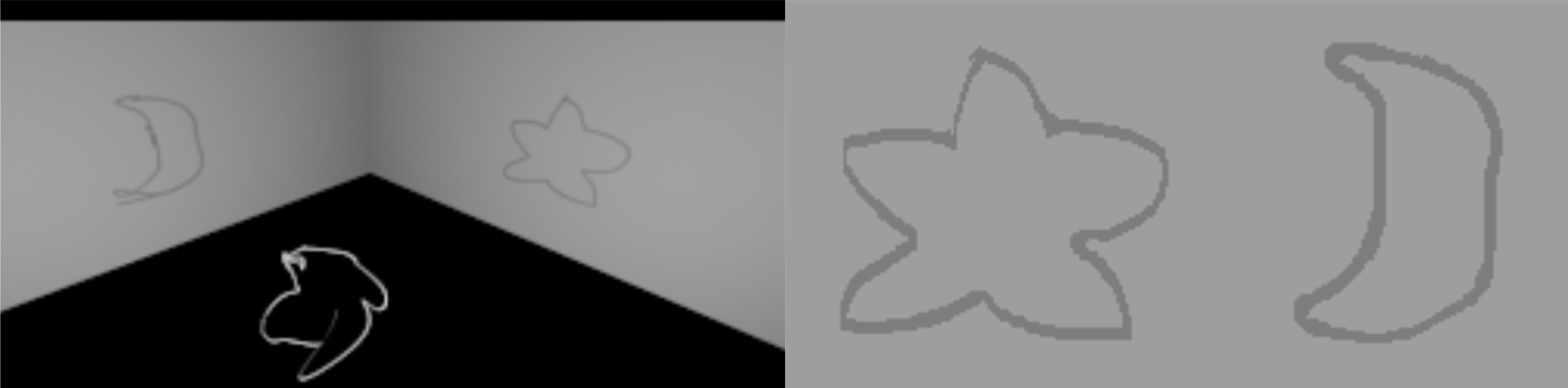}}
\caption{\textit{Mitsuba}'s render and shadow \cite{b1}.}
\label{fig}
\end{figure}

This project seeks to achieve the same goal, but with a much simpler model that can accurately make objects to match Kanji (Chinese characters adapted into the Japanese language) as shadows. It does not utilize any pre-existing inverse rendering applications nor integration techniques.

\subsection{Motivation for Finding an Alternative to Projective Sampling Integrators}
When attempting to run \cite{b1}'s code on Kanji as inputs, the model fails miserably. This results from the fact that most Kanji are not topologically equivalent to a circle, which is imperative as the existing model runs on the assumption that a manifold curve can be molded in such a way where it can project the target shadows. Even beyond this major logistical discrepancy, the model utilizes multiple-gradient descent, which makes it overly slow and unpredictable. 

Most target shapes--- not just Kanji--- are thus not suitable for this method, and none are optimized, highlighting a need for a new approach.
To solve this problem, this project uses voxels as a discrete approximation to a curve's continuous properties. For the purpose of printability, each voxel only has two properties: whether or not it exists and its relative position.

\section{Assumptions and Tools}
\subsection{Formal Mathematical Assumptions}
We model the voxels as a three-dimensional binary cubic tensor, $T \in \{0, 1\}^{n \times  n \times n}$ of side length $n \in \mathbb{Z}^+$. Each value in the tensor represents a voxel with a relative physical position. A value of $1$ denotes an existent voxel, while a value of $0$ denotes a non-existent voxel.

The two models take in one, two, or three binary bitmaps as the target shadows. In this way, the models are called 1-way, 2-way, or 3-way projections. The symbols $P, Q, R$ are overloaded and used in two distinct contexts.

\begin{enumerate}
    \item \textbf{In an algebraic context,} $P, Q, R \in \{0, 1\}^{n \times n}$ represent matrices with binary entries, indicating the presence (1) or absence (0) of specific features in a 2D plane projection.

    \begin{itemize}
         \item Assume each combination of pixels from  the three planes (or, if $Q$ or $R$ do not exist, then each combination of hypothetical pixels if they did exist) corresponds to exactly one voxel in $T$.  
         \item Suppose that, when specified, the matrices $P_\text{in}, Q_\text{in}, R_\text{in}$ refer to the inputted target shadows and that the matrices $P_\text{out}, Q_\text{out}, T_\text{out}$ refer to the outputted resultant shadows of the optimized $T$.
    \end{itemize}
    
    \item \textbf{In a geometric context,} $P, Q, R$ are finite planes in $\mathbb{R}^3$ space, representing the projections of the 3D object along different axes.

    \begin{itemize}
        \item The standard basis of this real vector space is $\{\hat{i}, \hat{j}, \hat{k}\} \subset \mathbb{R}^3$. For consistency, assume that $P \perp \hat{i}$, $Q \perp \hat{j}$, and $R \perp \hat{k}$.
        \item $P, Q, R$ are each orthogonal to each other and to a side of the cube representing the tensor, which can also be geometrically thought of as a cube, $T$. Each plane perfectly aligns with one of the cube's six faces.
        \item Suppose $(\psi, \theta, \phi)$ represent the rotational position of $T$.
    \end{itemize}
\end{enumerate}

In both cases, $P$ is a mandatory hyperparameter, $Q$ is optional, and $R$ is optional but can only exist if $Q$ exists.

Finally, assume that there are exactly three planar light sources, each infinitely bright, infinitely far away, pointing towards the cube $T$, and each normal to each of the three canonical vectors $\hat{i}, \hat{j}, \hat{k}$. For simplicity and to assert Monte Carlo integration is not necessary, assume the objects cast perfect shadows.

Assume that light rays are pixels that travel perfectly straight and are parallel to both their light source and target plane, and assume they do not interact with other light pixels. Note the notation $[n] = \{1, \cdots, n\}$.

\subsection{Technological Specifications}
The code is written entirely in Python \texttt{3.8.10} and NumPy \texttt{1.24.3}. SciPy \texttt{1.10.1} is used exclusively for solving the linear program using the \texttt{scipy.optimize.linprog} function. Please see the Appendix for the project's accompanying code. The computations were performed using the Computer Vision Laboratory's GPU servers at Osaka University| Suita campus; it was specifically a system equipped with 10 Tesla V100 GPUs, CUDA 12.1, and 768GB RAM, running on an Intel Xeon Gold 6142 CPU @2.60GHz under Ubuntu 20.04.1 LTS. However, the computations are efficient enough to run well on everyday computers. It goes without saying that the 3D printing portion of this project required the use of the laboratory's 3D printer, which was a Raise3D Pro3 Plus at the time of this project.

In order to process the data for this project, there is a multi-step data pipeline. First, the character is inputted as a character (\texttt{str} in Python). Then, it is loaded into a two-dimensional bitmap represented by a NumPy array, then converted to an alpha channel, then a mesh (with vertices and faces), and finally a \text{.obj} file which can then be remeshed in software like Blender.

\section{Linear Programming Model}

A linear program (abv. LP) is a model that optimizes $\vec{x}$ in the set of inequalities

$$\max_{\vec{x}} \vec{c}^{~T} \vec{x} \text{ s.t. } \begin{cases}
    A \vec{x} \leq \vec{b}\\
    \vec{x} \geq 0
\end{cases} \text{\cite{b4}.}$$

In other words, it finds the optimal values of $\vec{x} \geq 0$ from a linear sum subject to a set of linear inequalities. We use a linear program to solve for our voxels.

\subsection{Setting up the Linear Program}
In order to construct a linear program that can solve the problem, we use the entries in the tensor as the set of variables to optimize. So, we inject the tensor into the (very large) reshaped L.P. vector, $\vec{x} \in \mathbb{R}^{n^3}$. Note that the injection function, $F$, needs to be a bijection (i.e. an isomorphism) so that the vector can be reshaped back into the tensor. To note the abuse of notation, let $x_{i,j,k} \in \vec{x}$ represent ``the element of vector $\vec{x}$ corresponding to the element of the tensor $T_{i,j,k}$" and let $\vec{x}_\text{bin}$ represent a binary version of $\vec{x}$. We therefore have
$$F: T \to \vec{x}$$
and
$$F^{-1} : \vec{x} \to T \text{.}$$

Although a binary $\vec{x}_\text{bin} \in \{0, 1\}^{n^3}$ would be more suitable to the problem, we unfortunately cannot do this as Zero-One Equation (ZOE) linear programming is well known to be NP-Complete \cite{b5}. We can, however, approximate the problem by relaxing our constraints, allowing $\vec{x}$ to be a continuous vector between the values of $0$ and $1$. 

This approximation makes an obvious violation of our core assumption that each vector either exists or does not exist (without existing in some quantum superposition between the two), which we can deal with in one of two ways. First, we  could just accept the fact that $T$ will (in all likelihood) contain non-binary values. One way this could work is by having each value $\alpha = 255x_{i,j,k}$ represent a transparency ($0 \leq \alpha \leq 255$) value of the output mesh, however, this model cannot be three-dimensionally printed unless given access to very specific printer technology capable of supporting a wide range of varying material properties. The second way to deal with this problem, which this project ultimately employs, is to postcompute binary values, $\vec{x}_\text{bin}$, after the linear program has been solved. This project arbitrarily chose a global threshold, $0 < \lambda < 1$, to calculate each value of $\vec{x}_\text{bin}$, we have
$$b = \begin{cases}
    0, & x < \lambda \\
    1, & x \geq \lambda
\end{cases} ~~\forall b \in \vec{x}_\text{bin}, ~ x \in \vec{x} \text{.}$$
In hindsight, 
having such a global threshold discriminator produces resultant projection shadows that are highly similar to the target. In fact, when the finalized linear program was tested with an allowable error for each inequality as $\epsilon > 0$, defining the global threshold as anywhere within the interval $\epsilon < \lambda < 1$ consistently produces perfect or near-perfect results.

\begin{figure}[htbp]
\centerline{\includegraphics[width=250px]{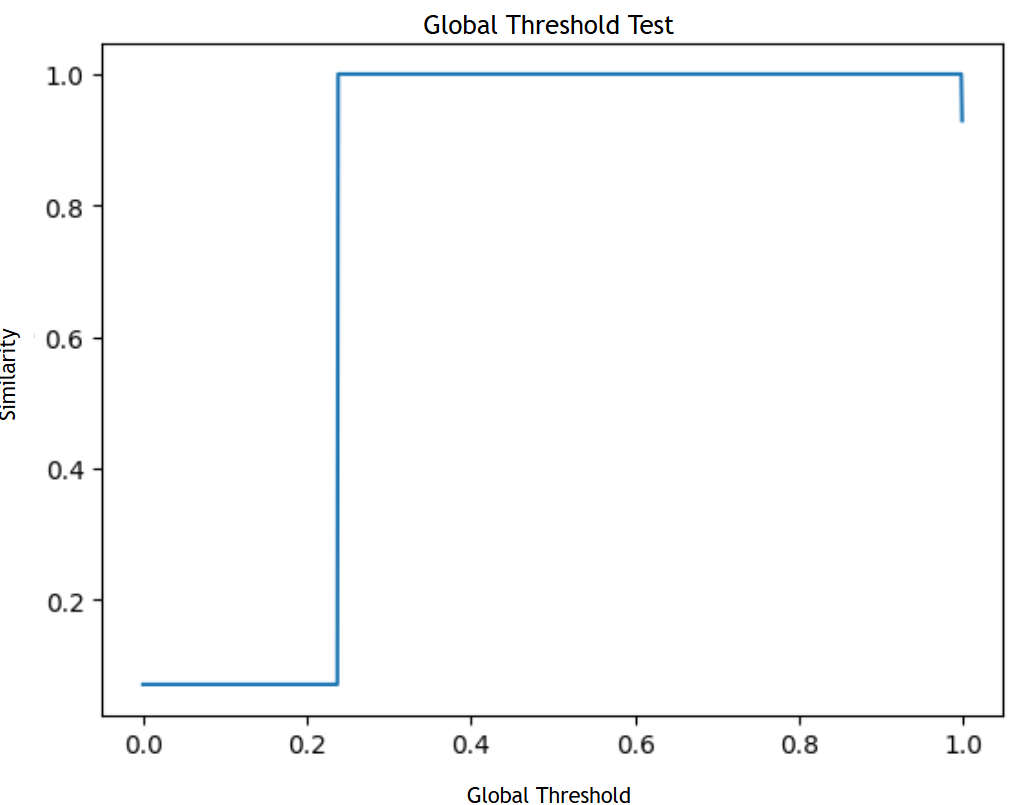}}
\caption{Graph showing the relationship between the global threshold 
λ and the \textit{similarity percentage} (the fraction of projection pixels matching the target projection) when $\epsilon = 0.239$.  The graph indicates a sharp increase in similarity at $\lambda \approx 0.239$. Indeed, additional tests indicate that an optimal threshold value of $\lambda = \underset{\delta \to 0^+}{\lim} \delta + \epsilon$.}
\label{fig}
\end{figure}

\subsection{Constructing A Linear Programming Model}

To begin constructing the linear program, begin by considering the constraints. We start with the the trivial constraint that each value must have a valid state.

\begin{equation}
    0 \leq x_{i,j,k} \leq 1 ~~~~ \forall x_{i,j,k} \in \vec{x} \text{.}
\end{equation}

For the rest of this construction, ignore the fact that the assumption that $x_{i,j,k} \in \{0, 1\} ~\forall x_{i,j,k} \in \vec{x}$ is contradictory to (1) because the model works very accurately despite this (its main problem is actually runtime, which is not caused by this assumption).

Notice that all three planes $P, Q, R$ are symmetrical to the unoriented cube $T$, as each plane is orthogonal to one of and parallel to the other two of the three standard basis vectors (e.g. $P \perp \hat{i}$ and $P \parallel \hat{j}, \hat{k}$). Making the additional assumption that any positive value of $T$ blocks light, we need to assert that each nonexistent pixel is sufficient and each existent pixel is necessary.

The former is achieved by setting constraints

$$
    \sum_{i \in [n]} x_{i,j,k} \leq \lim_{C \to \infty} CP_{j,k} ~~~~ \forall j,k \in [n] \text{,}
$$
$$
    \sum_{j \in [n]} x_{i,j,k} \leq \lim_{C \to \infty} CQ_{i,k} ~~~~ \forall i,k \in [n] \text{,}
$$

and

$$
\sum_{k \in [n]} x_{i,j,k} \leq \lim_{C \to \infty} CR_{i,j} ~~~~ \forall i,j \in [n] \text{,}
$$

which can be relaxed to
\begin{equation}
    \sum_{i \in [n]} x_{i,j,k} \leq nP_{j,k} ~~~~ \forall j,k \in [n] \text{,}
\end{equation}
\begin{equation}
    \sum_{j \in [n]} x_{i,j,k} \leq nQ_{i,k} ~~~~ \forall i,k \in [n] \text{,}
\end{equation}

and

\begin{equation}
    \sum_{k \in [n]} x_{i,j,k} \leq nR_{i,j} ~~~~ \forall i,j \in [n]
\end{equation}

because the sums cannot possibly exceed $n$.

The latter is achieved by setting constraints

$$
    \sum_{i \in [n]} x_{i,j,k} \geq \lim_{\delta \to 0^+} \delta P_{j,k} ~~~~ \forall j,k \in [n] \text{,}
$$
$$
    \sum_{j \in [n]} x_{i,j,k} \geq \lim_{\delta \to 0^+} \delta Q_{i,k} ~~~~ \forall i,k \in [n] \text{,}
$$

and

$$
\sum_{k \in [n]} x_{i,j,k} \geq \lim_{\delta \to  0^+} \delta R_{i,j} ~~~~ \forall i,j \in [n] \text{,}
$$

which can be relaxed to
\begin{equation}
    \sum_{i \in [n]} x_{i,j,k} \geq P_{j,k} ~~~~ \forall j,k \in [n] \text{,}
\end{equation}
\begin{equation}
    \sum_{j \in [n]} x_{i,j,k} \geq Q_{i,k} ~~~~ \forall i,k \in [n] \text{,}
\end{equation}

and

\begin{equation}
    \sum_{k \in [n]} x_{i,j,k} \geq R_{i,j} ~~~~ \forall i,j \in [n]
\end{equation}

from the assumption that the pixels are binary (simply ignoring the fact that this is a relaxed constraint). 

Finally, we must choose a linear optimization problem. To maximize three-dimensional printability, this project chooses a maximization object, and we arbitrarily decide to have no particular weights.
Thus, our final linear program is

$$\max \sum_{i,j,k \in [n]} x_{i,j,k} \text{ s.t. } \begin{cases}
    0 \leq x_{i,j,k} \leq 1 & \forall i,j,k \in [n] \hfill (1) \vspace{2.5mm}\\

        \underset{i \in [n]}{\sum} x_{i,j,k} \leq nP_{j,k} & \forall j,k \in [n] \hfill (2)\\
        \underset{j \in [n]}{\sum} x_{i,j,k} \leq nQ_{i,k} & \forall i,k \in [n] \hfill (3)\\
        \underset{k \in [n]}{\sum} x_{i,j,k} \leq nQ_{i,j} & \forall i,j \in [n] \hfill (4)\\
        \underset{i \in [n]}{\sum} x_{i,j,k} \geq P_{j,k} & \forall j,k \in [n]  \hfill (5) \\
        \underset{j \in [n]}{\sum} x_{i,j,k} \geq Q_{i,k} & \forall i,k \in [n] \hfill (6) \\
        \underset{k \in [n]}{\sum} x_{i,j,k} \geq R_{i,j} & \forall i,j \in [n] \hfill ~~~~(7)
\end{cases} \text{,}$$

with a constant allowable error $\epsilon > 0$. Now, we have solved for the values of the tensor, and implementation is just a matter of appropriately using existing LP software.

\subsection{Results Anti-Aliasing}

Even after postcomputation with the global threshold $\lambda$, in order for the model to work in a reasonable time, the resolution must be very low, which can cause severe aliasing of the output mesh and its projected shadows.

\begin{figure}[htbp]
\centerline{\includegraphics[width=250px]{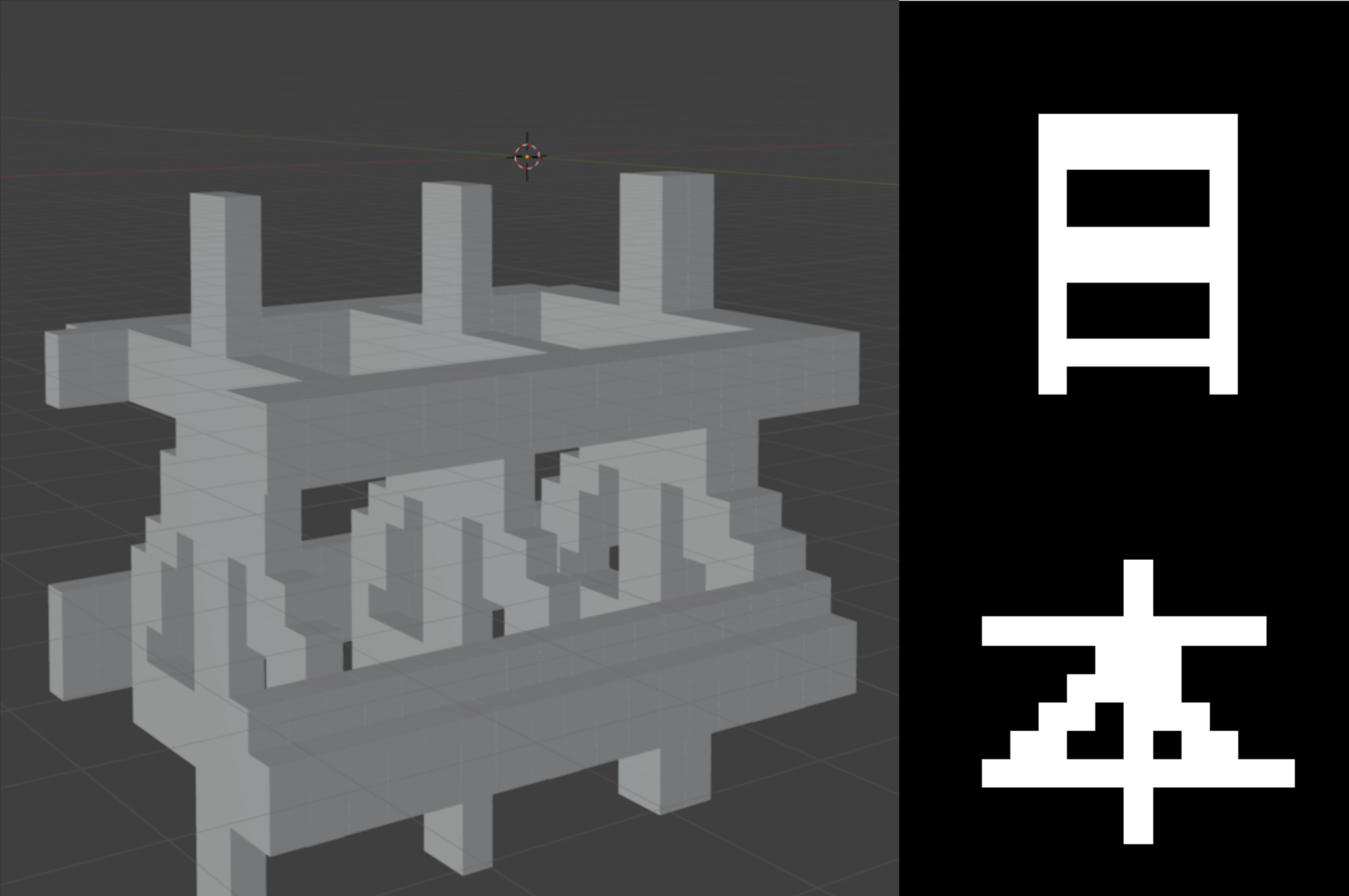}}
\caption{This rendered mesh demonstrates the current model's aliasing problem.}
\label{fig}
\end{figure}

We can still edge out shadows of even higher resolution. Traditional two-dimensional anti-aliasing techniques for $n \times n$ resolution image take an $N \times N$ supersample of that image where $N \in \mathbb{Z}^+$ is some anti-aliasing factor. Then, the supersamples are aggregated (typically averaged out) in the final frame buffer \cite{b6}.

\begin{figure}[htbp]
\centerline{\includegraphics[width=250px]{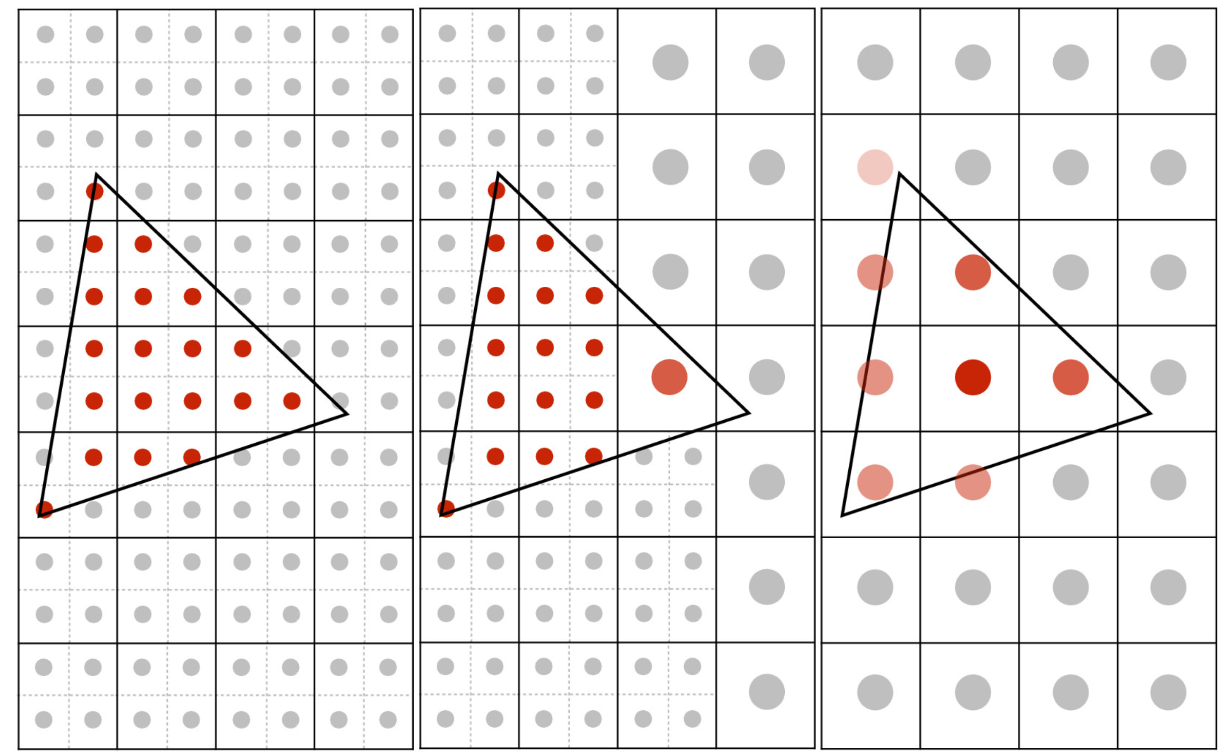}}
\caption{Traditional anti-aliasing by supersampling \cite{b6}.}
\label{fig}
\end{figure}

According to the Nyquist Theorem, in order to achieve no aliasing, the frequency $f$ of the signal must be less than the Nyquist frequency, in other words,
$$f < f_\text{Nyquist} := \frac{1}{2} f_\text{sample} \text{\cite{b7}.}$$
Supersampling with $N = 8$ ensures that the sampling frequency is sufficiently high to meet the Nyquist criterion, thereby minimizing aliasing artifacts while balancing computational complexity and image quality. Traditional anti-aliasing contradicts the model's assumption that voxels take on binary values, so the project instead employs a supersampled carving technique. Note this means that the total number of supersampled points for each dimension is the product $Nn$, because there are $n$ samples each with $N$ supersamples.
 
Let matrices $P_\text{sup}, Q_\text{sup}, R_\text{sup} \in \mathbb{R}^{Nn \times Nn}$ denote the  supersampled target projections corresponding to $P_\text{in}, Q_\text{in}, R_\text{in}$ respectively. Let us also dialate the tensor $T \in \mathbb{R}^{n \times n \times n}$ to the supersampleable tensor $T_\text{sup} \in \mathbb{R}^{Nn \times Nn \times Nn}$.

For each existent supersampled subvoxel $(T_\text{sup})_{i,j,k} = 1$, if a subpixel is inexistent in all three corresponding supersampled subpixels, $(P_\text{sup})_{j, k} = (Q_\text{sup})_{i, k} = (R_\text{sup})_{i, j} = 0$, then the existence of this subvoxel necessarily makes the image quality worse. So, we can just delete it; let us call this method "carving." We  have thus reached the anti-aliasing boolean expression
$$(T_\text{sup})_{i,j,k} \gets \lnot (P_\text{sup})_{j,k} \land \lnot (Q_\text{sup})_{i,k} \land \lnot (R_\text{sup})_{i,j} \text{.}$$

Implementing this into code achieves its desired result.

\begin{figure}[htbp]
\centerline{\includegraphics[width=250px]{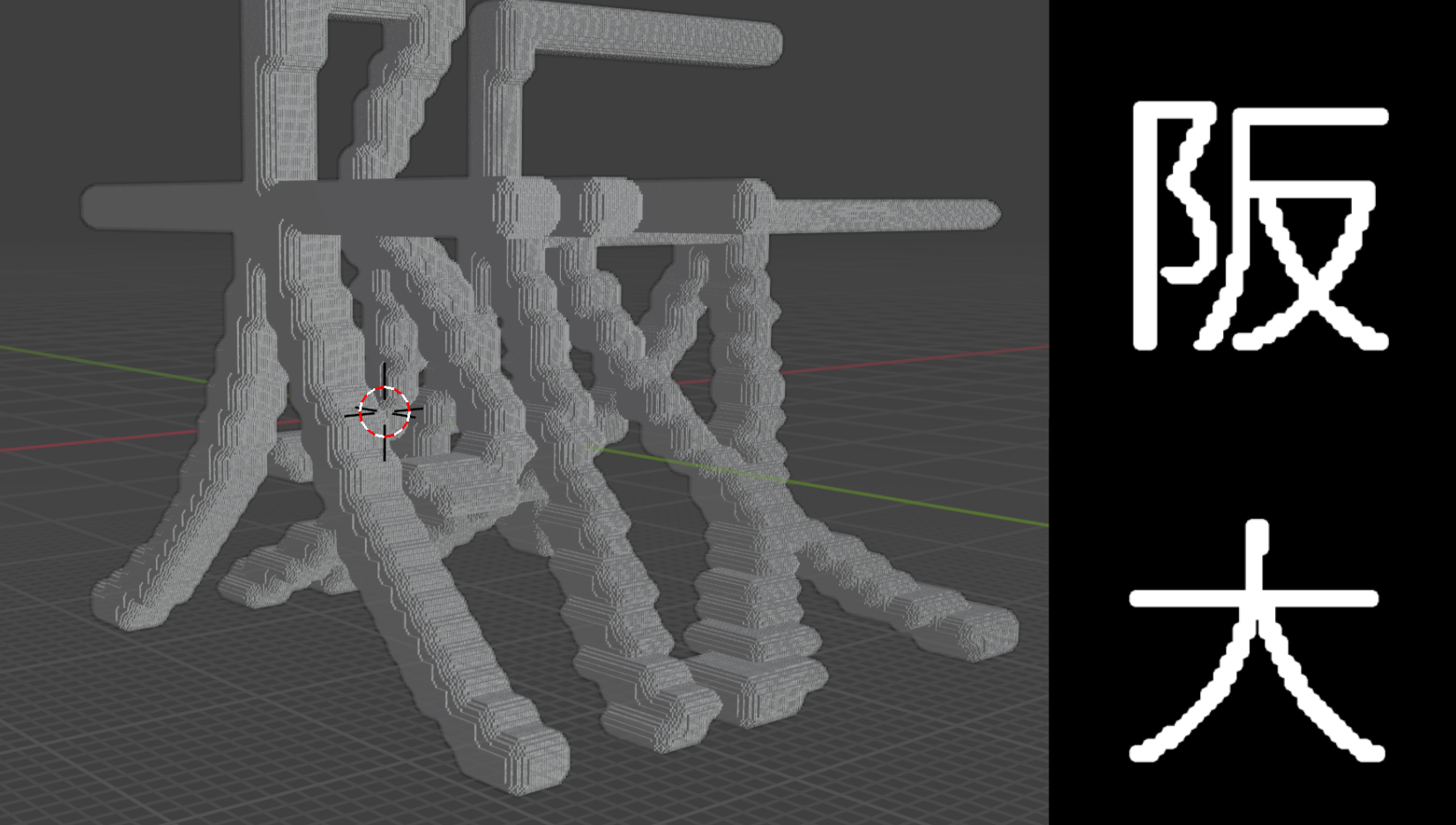}}
\caption{A rendered, anti-aliased mesh with $(n, N) = (32, 8)$.}
\label{fig}
\end{figure}

\subsection{Theoretical Runtime (In)efficiency}
Let $t_\text{LP}(n)$ denote the runtime of this model given $n$, the model's aliased resolution. The runtime of solving a linear program using the traditional simplex algorithm is $t_\text{LP}(n) = O(m_1(m_1 + m_2))$ where $m_1$ is the number of optimizing variables and $m_2$ is the number of inequality constraints \cite{b4}.

Calculating $m_1$ is trivial and given by

$$m_1 = \text{len}(\vec{x}) = n^3 \text{.}$$

Furthermore, $m_2$ can be calculated through discrete set counting. Note that, because the tensor $T$ is cubic, each of the three sets of sums (2), (3), (4) have the same size. We have

$$m_2 = |\{ (i,j,k)~|~\forall i,j,k \in [n]\}| + 3|\{ (j, k)~|~\forall j, k \in [n] \}| $$
$$= |[n] \times [n] \times [n]| + 3|[n] \times [n]|$$
$$= n^3 + 3n^2 \text{.}$$

Plugging in the values for $m_1$ and $m_2$, we have thus obtained the theoretical runtime of this model, which is

$$t_\text{LP}(n) = O(n^3(n^3 + n^3 + 3n^2)) = O(n^6) \text{.}$$

Note that, assuming $n \gg N$, the addition of the anti-aliasing feature does not increase asymptotic runtime as it is dominated by the linear program solver.

\subsection{Empirical Runtime (In)efficiency}

This project utilizes the High Performance Optimization Software (HiGHS) in place of the simplex algorithm for best performance \cite{b8}. This allowed the model to achieve the tested empirical runtime, much better than was theoretically expected,

$$t_{LP}(n) \approx 0 \pm n^{4.022} \text{~(fig. 8).}$$

When the traditional simplex algorithm was tested, the model performed much closer to the theoretical worst-case runtime $O(n^6)$.

\section{Direct Carving Model}
\subsection{Motivation}
Even with the HiGHS runtime acceleration, the linear programming model is still painfully slow. Furthermore, even with anti-aliasing, the use of the low-resolution linear programming model makes the output render have large, strange ``bubbles." Instead of optimizing the program at a lower resolution and anti-aliasing at a higher resolution, the direct carving model starts with an all $1$'s tensor, $T_\text{in} = \mathds{1}_{n \times n \times n}$, and carves it directly without solving a linear program.

\subsection{Algorithm}
The algorithm is surprisingly simple and elegant. The following algorithmic description is for 3-way projections. The way it works is that, for a of the three planes, for all $0$ pixels in that plane, delete all tensor voxels orthogonal to that pixel. For style, the existence of $R$ necessarily assumes the existence of $Q$.

\begin{figure}[htbp]
\centerline{\includegraphics[width=250px]{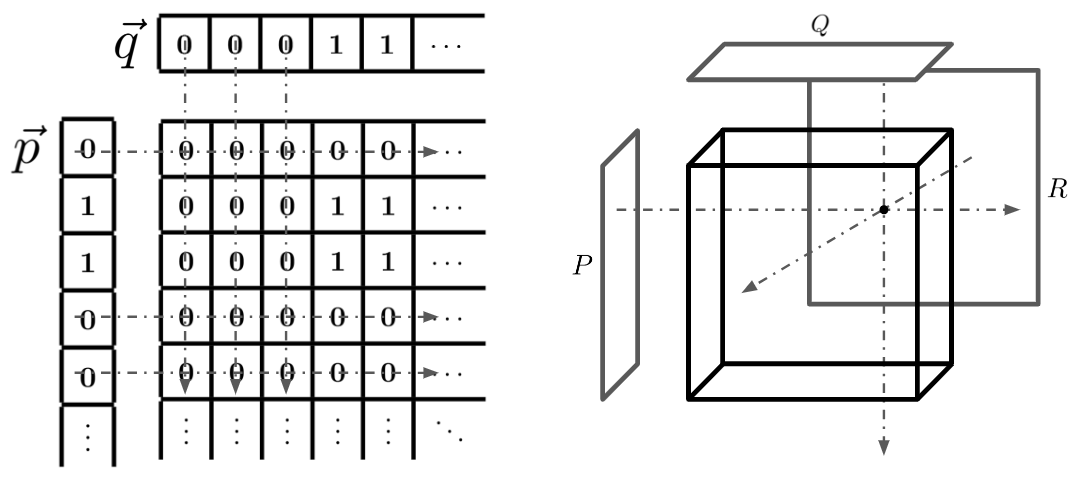}}
\caption{Visualization of the direct carving model in 2D and 3D space.}
\label{fig}
\end{figure}

The resultant meshes not only succeed in matching the target shadows more often than the L.P., but they look much better, too. 

\begin{figure}[htbp]
\centerline{\includegraphics[width=250px]{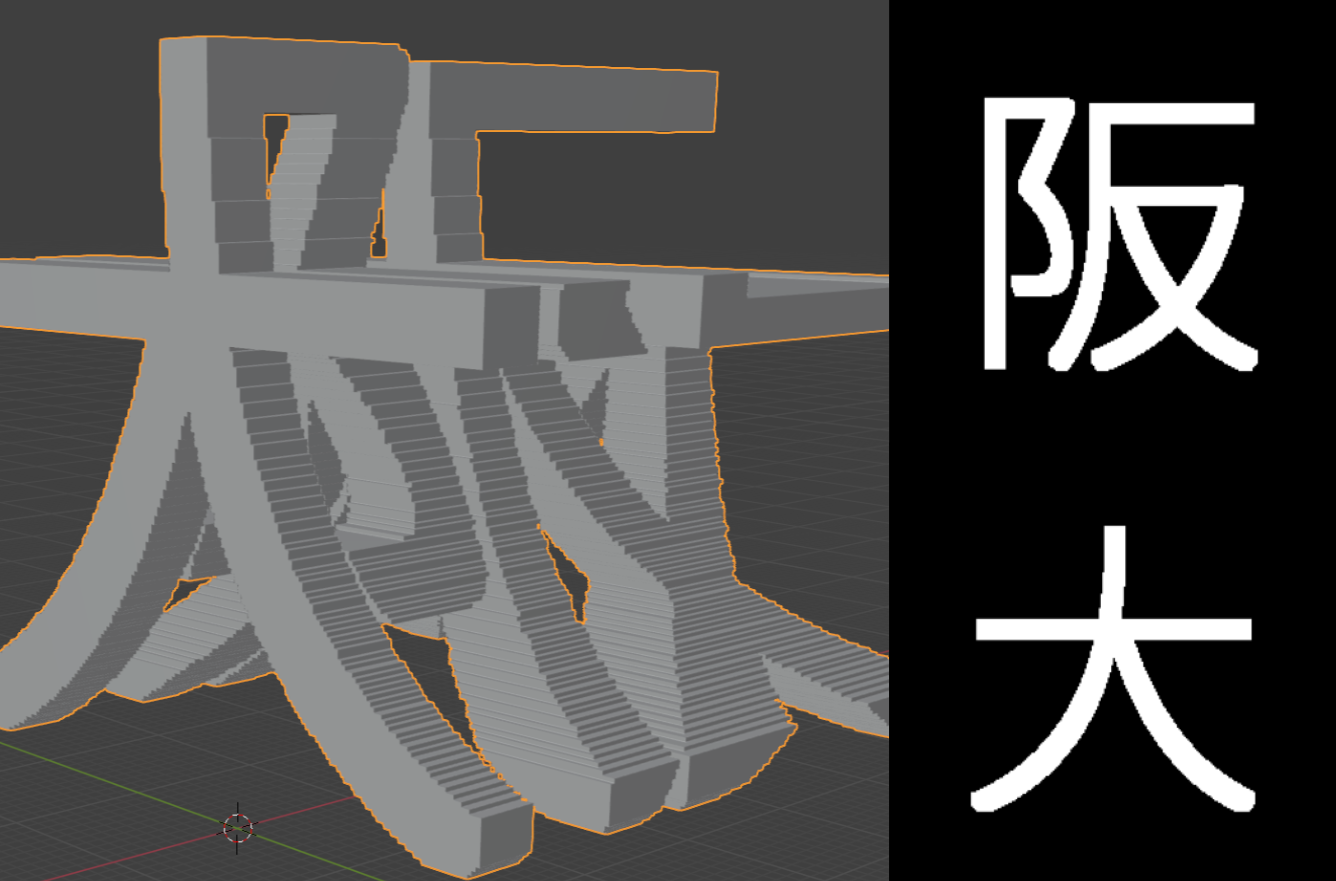}}
\caption{Directly carved mesh| a much higher resolution is possible with the same runtime.}
\label{fig}
\end{figure}

\subsection{Theoretical Runtime Efficiency}
In the worst case (i.e. assuming single-threaded programming and no optimization), the instantiation of the tensor, $T \gets \mathds{1}_{n \times n \times n}$, takes $O(n^3)$ time and the three for-loops each take $O(n^2)$ time. We thus have the runtime $t_\text{Carve}(n)$ of the algorithm,
$$t_\text{Carve}(n) = O(n^3) + O(n^2) = O(n^3) \text{.}$$

\subsection{Empirical Runtime Efficiency}
When tested on three target projections, the algorithm empirically performed better than expected,

$$t_\text{Carve}(n) \approx 0 \pm n^{2.990} \text{~(fig. 8).}$$

This suggests that there are a number of NumPy optimizations that benefit the runtime of this algorithm. This further prompts future work in finding the best optimization scheme for this algorithm.

\subsection{Abstraction to Higher Dimensions}

Besides the physical limitations of the universe, there is no reason to stop at three dimensions-- as the direct carving model can work in 2-dimensional (as alluded to in fig. 6) and arbitrarily high dimensional binary space. As suggested in fig. 8, the direct carving model easily works for projecting a 2-dimensional matrix $M \in \{0, 1\}^{n \times n}$ on (at most) two target 1-dimensional vectors $\vec{p}, \vec{q} \in \{0, 1\}^n$. Indeed, this model is abstractable to any number of higher dimensions.

Let $d \in \mathbb{Z}_{\geq 2}$ be the dimensionality of the output hypertensor and let $d^\prime \in \mathbb{Z}^+ : d^\prime \leq d - 1$ be the number of hyperplane images. Let $P_1, \cdots, P_{d^\prime} \in \{0, 1\}^{n^{d - 1}}$ be the aforementioned hyperplane target images. Let $H \in \{0, 1\}^{n^d}$ be the hypertensor output.

As before, we carve along the $n^{d - 1}$ hypervoxel normals from each of the $d^\prime$ hyperplanes facing the hypertensor. Note this cannot be visualized and the use case is currently unclear.

As for runtime, the creation of the hypertensor takes $O(n^d)$ time and carving with the $d^\prime$ hyperplane target images takes $d^\prime \times O(n^{d-1})$ time. Because, by definition, $d^\prime < d$, we have the theoretical runtime

$$t_\text{Carve}(n, d) = O(n^d) + d^\prime \times O(n^{d-1}) = O(n^d) \text{.}$$

\section{Model Comparison}

\begin{table}[htbp]
\caption{Runtime Summary}
\begin{center}
\begin{tabular}{|c||c|c|}
\hline
& \textbf{Theoretical} & \textbf{Empirical}\\
\hline
\hline
\textbf{Linear Programming Model} & $O(n^6)$ & $0 \pm n^{4.022}$ \\
\hline
\textbf{Direct Carving Model} & $O(n^3)$ & $0 \pm n^{2.990}$\\
\hline
\end{tabular}
\end{center}
\end{table}

\begin{table}[htbp]
\caption{Emperical Runtime}
\begin{center}
\begin{tabular}{|c||c|c|c|c|}
\hline
& $n = 64$ & $n = 256$ & $n = 1028$ & $n = 4096$\\
\hline
\hline
$t_\text{LP}$ & 49 seconds & 3.6 hours & 40 days & 28.6 years\\
\hline
$t_\text{Carve}$ & 0.22 seconds & 14 seconds & 14.5 minutes & 15 hours\\
\hline
\end{tabular}
\end{center}
\end{table}

\begin{figure}[htbp]
\centerline{\includegraphics[width=250px]{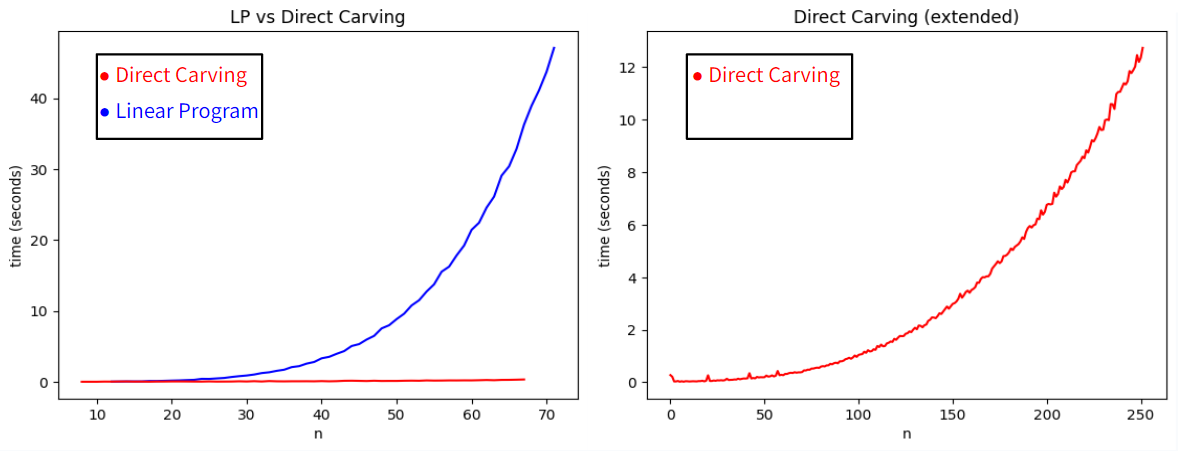}}
\caption{Runtime comparison of the L.P. model versus the Direct Carving model. As expected, the linear programming model is polynomially slower than the direct carving model. Also as expected, the direct carving model itself follows a polynomial runtime.}
\label{fig}
\end{figure}

Both the theoretical and empirical results show that, in terms of runtime, direct carving is the superior model to linear programming.


\section{Three-Dimensional Printing}
The results of 3-dimensional printing on the correctly adjusted settings came out wonderfully. 

\vspace{5mm}

\begin{figure}
{\footnotesize
    \begin{tabular}{ccc} 
        \includegraphics[height=70px, width=70px]{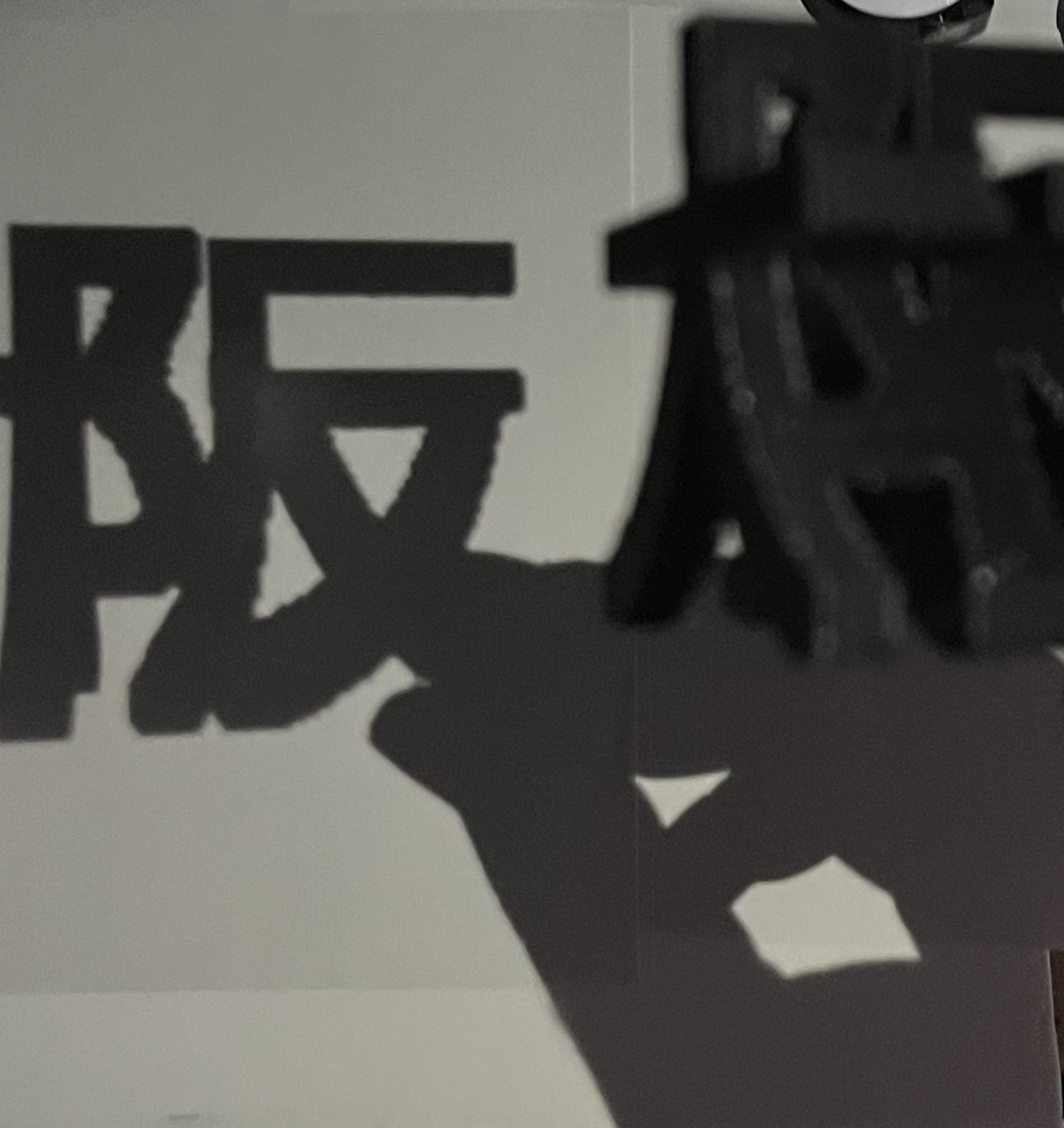} & 
        \includegraphics[height=70px, width=70px]{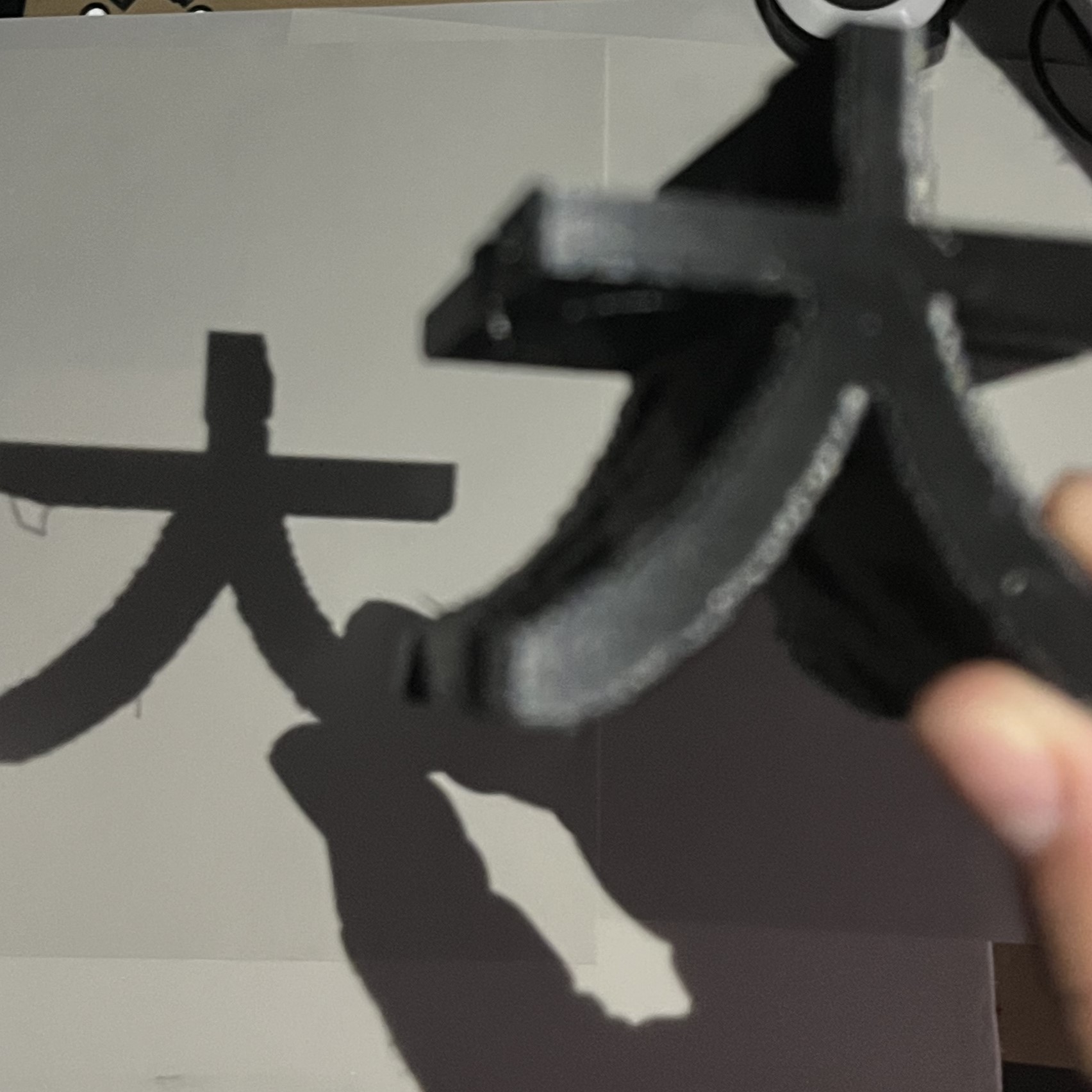} &\\
        ``Handai'' $(0^\circ, 0^\circ, 0^\circ)$ & 
        $(90^\circ, 0^\circ, 0^\circ)$ &    \end{tabular}

    \begin{tabular}{ccc}
            \includegraphics[height=70px, width=70px]{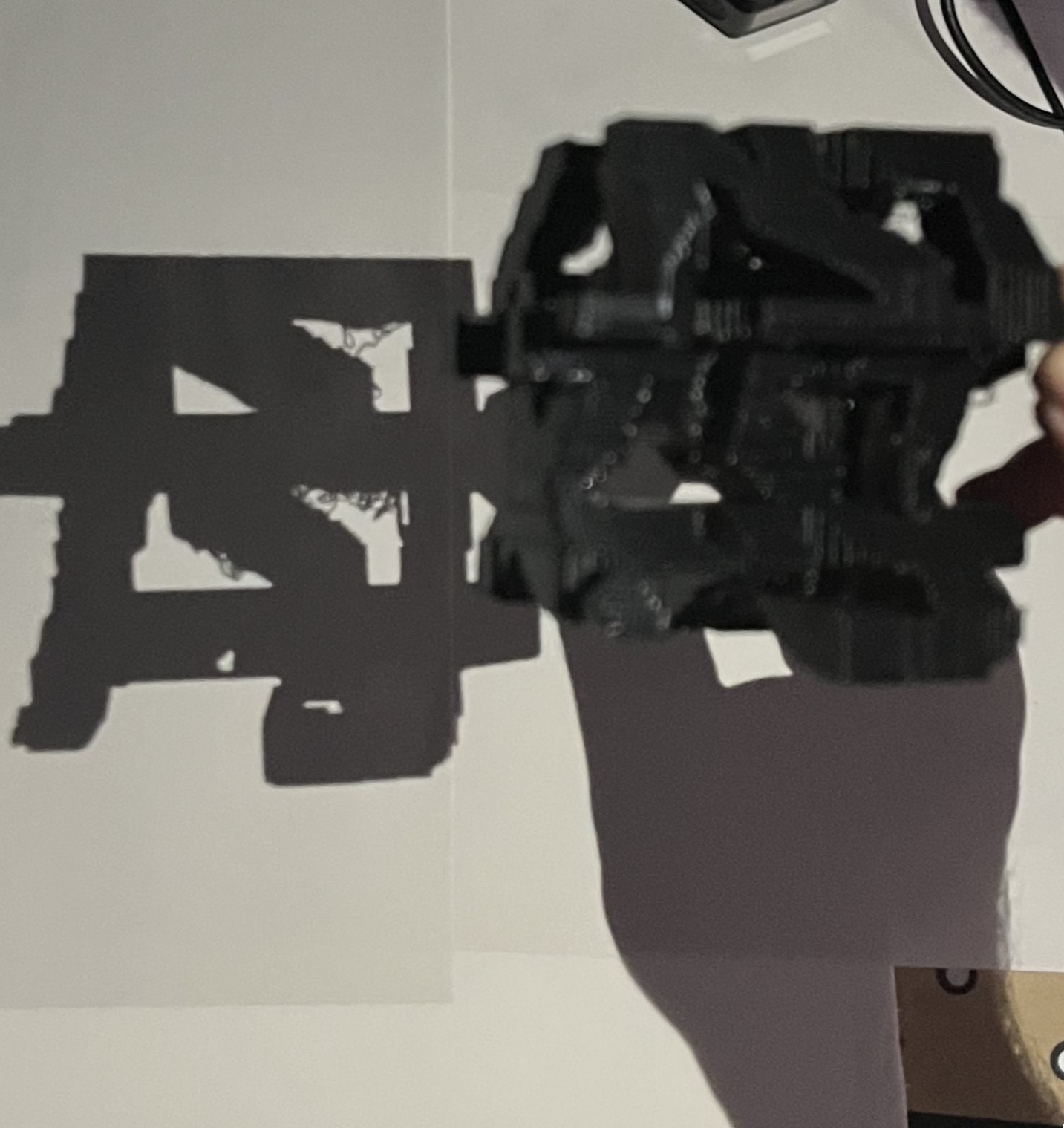} & 
            \includegraphics[height=70px, width=70px]{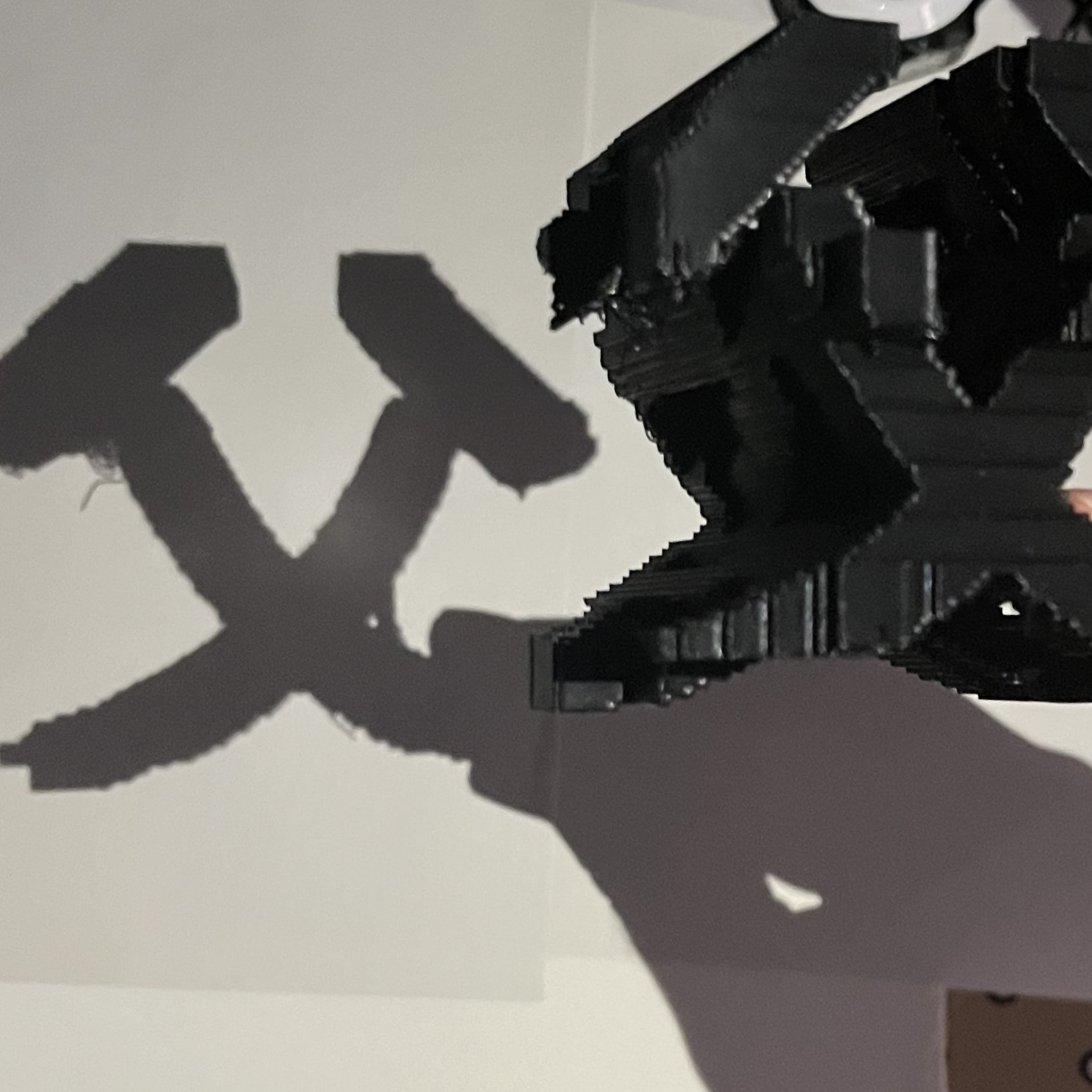} & 
            \includegraphics[height=70px, width=70px]{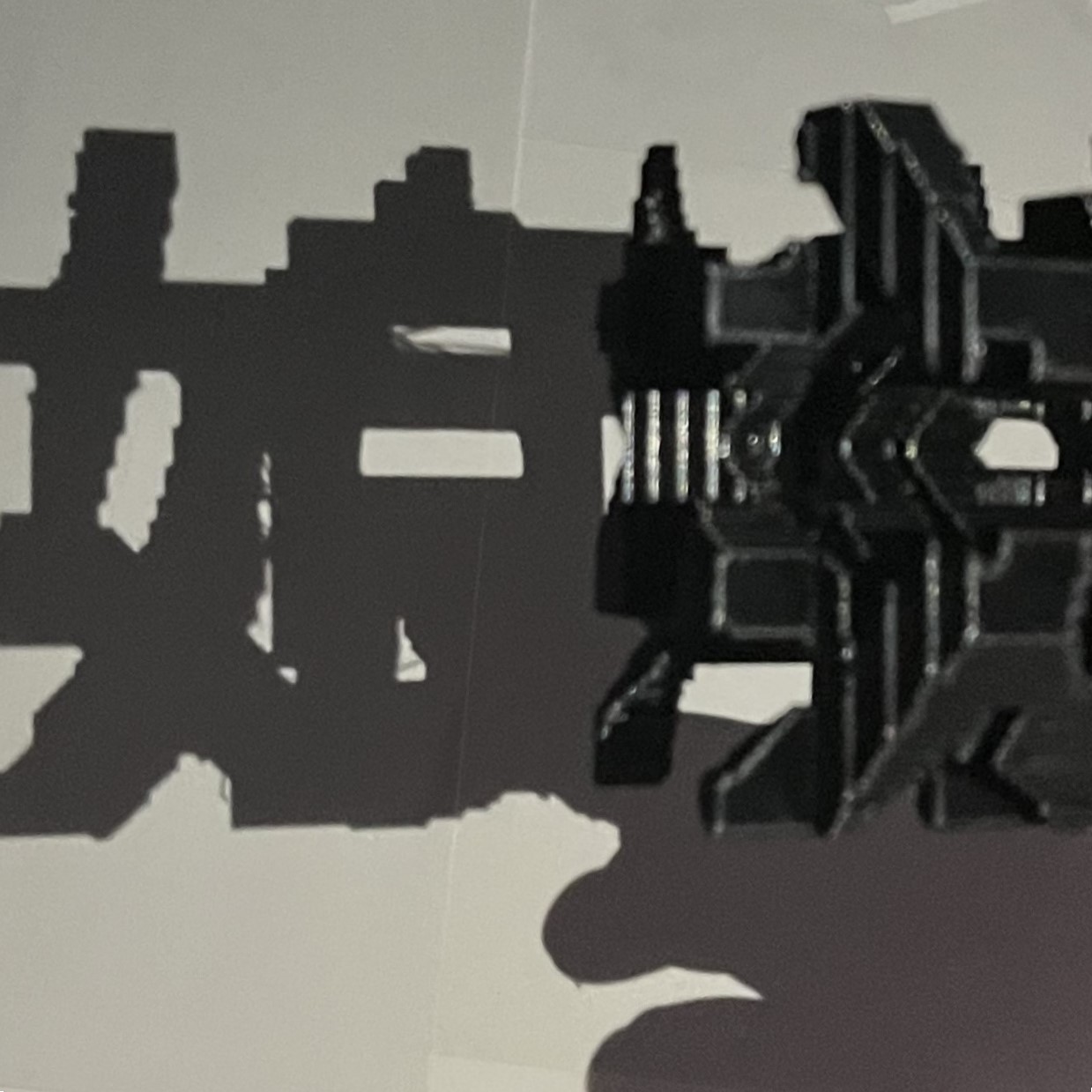} \\
        ``Kazoku'' $(0^\circ, 0^\circ, 0^\circ)$ & 
        $(90^\circ, 0^\circ, 0^\circ)$ &
        $(0^\circ, 0^\circ, -90^\circ)$
    \end{tabular}
}

\caption{Models printed with the laboratory's 3-dimensional printers.}
\label{fig}
\end{figure}

\begin{figure}[htbp]
\centerline{\includegraphics[width=250px]{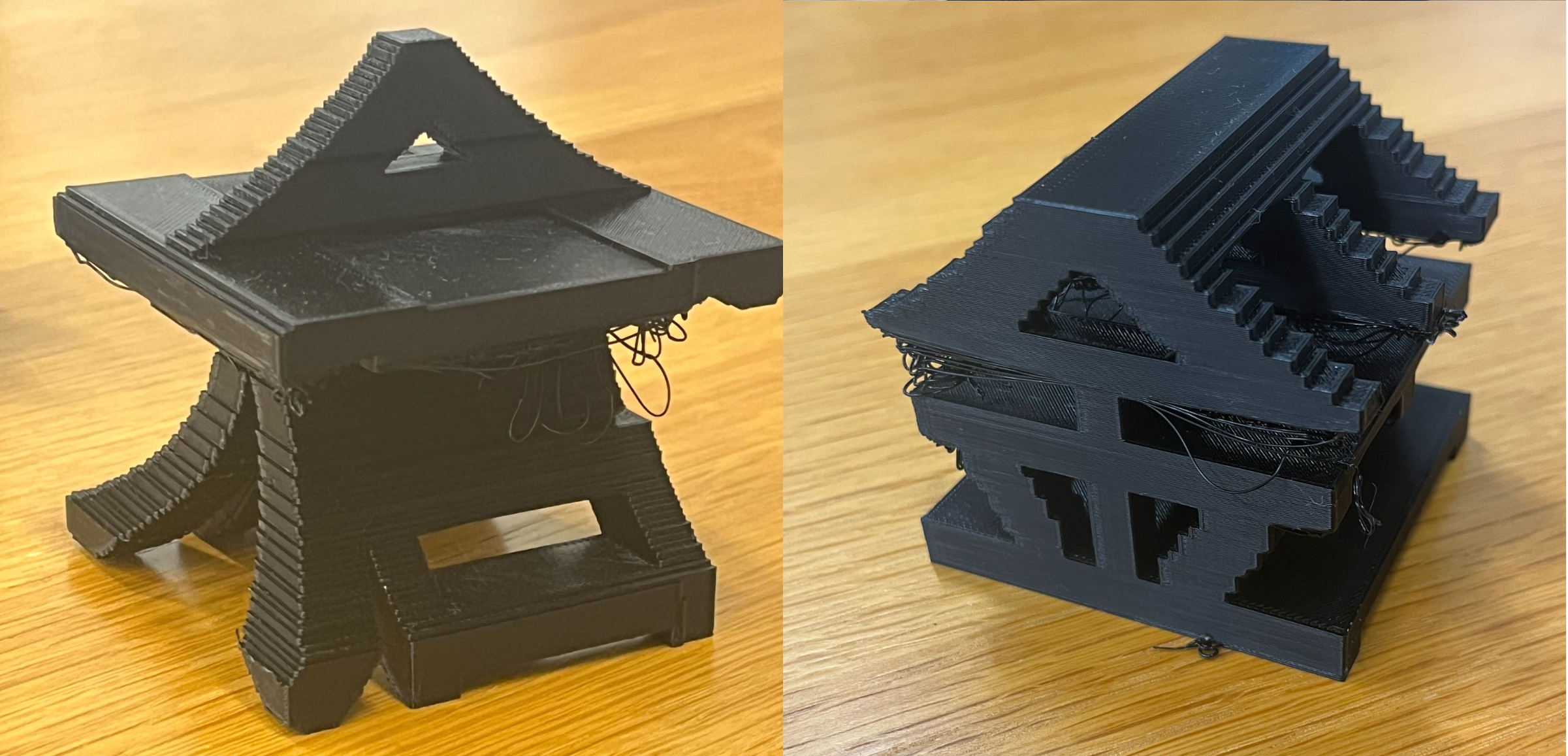}}
\caption{Kanji of  ``Ōkura" and ``Kaneda," two staff of the Matsushista Laboratory. Unfortunately, the Kanji for ``Matsushita" was not printable due to floating components.}
\label{fig}
\end{figure}

\section{Further Work}
The working time frame of this project was less than two months, leaving room for more than enough future work. 

Oblique projections are an interesting idea, and explored as a possibility for a use case of the linear program, but even that can be done with the direct carving mesh. 

Some renderable meshes are inevitably unprintable due to floating components. Still, they can be used for animation purposes. They can also be post-computed with additional components to ensure their printability. 

Finally, as alluded to in section III(a), we can specify an alpha value, $\alpha$, to represent a mesh's transparency, which can be three-dimensional printed using highly specialized technology. Going further, we could also implement RGB-shadowing capabilities to represent any shape of any color, which we could somehow print out using even more specialized technology. The possibilities are really as endless as the developer's imagination.

\section{Conclusion}
The initial goal of this project was to inversely render meshes with two different Kanji as shadows, similar to the approach used for curves by W. Jakob in his publications. To achieve this, three-dimensional continuous space was simplified into a binary tensor with one to three finite planes. Two methods were employed: one optimizing a set of linear, continuous inequalities, and the other a geometric, algorithmic approach. Of these, the latter performed significantly better. Not only was the initial goal completed, but the project also successfully works with one or three shadows, provides a clear pipeline for three-dimensional printing, and generalizes nicely to higher dimensions.


\vspace{12pt}

\section{Appendix}
Please see the following resources associated with this project.
\begin{itemize}
    \item \textbf{Repository:}\\
    https://github.com/wilrothman/sakira-code/tree/main\\
    \item \textbf{Code:}\\
    https://github.com/wilrothman/sakira-code/blob/main/sakira-code.py\\
    \item \textbf{FrontierLab@OsakaU Poster:}\\https://github.com/wilrothman/sakira-code/blob/main/SAKIRA\%20Poster.pdf\\
    \item \textbf{FrontierLab@Osaka Presentation:}\\https://github.com/wilrothman/sakira-code/blob/main/SAKIRA\%20Slides.pdf
\end{itemize}
\end{document}